%% file: lens-b.tex
\documentclass[11pt,fleqn]{article}

\usepackage{amsmath,amsfonts,amssymb,latexsym,cite}
\usepackage{graphicx}

\input preamble.tex

\def\rf{\eqref}
\def\eqn#1{Eq.\,\rf{#1}}
\def\MN{^{\mu\nu}}

\def\arccot{\mathop{\rm arccot}\nolimits}
\def\R{{\mathbb R}}

\def\wh{wormhole}
\def\whs{wormholes}
\def\bh{black hole}
\def\bhs{black holes}
\def\ssph{static, spherically symmetric}
\def\asflat{asymptotically flat}

\def\GR{general relativity}

\def\Swz{Schwarz\-schild}

\def\Figu#1#2#3{\begin{figure}
	\centering
	\includegraphics[width=#2]{#1} 
	\caption{\small #3}
		\end{figure} }
\usepackage{color}

\begin{document}
\twocolumn[
\thispagestyle{empty}
%\jnumber{1}{2019}

\Title{On gravitational lensing by symmetric and asymmetric wormholes}

\Aunames{K. A. Bronnikov\au{a,b,c,1} and K. A. Baleevskikh\au{a,2}} 

\Addresses{
\adr a {\small Center for Gravitation and Fundamental Metrology, VNIIMS,
             Ozyornaya ul. 46, Moscow 119361, Russia}
\adr b {\small Peoples' Friendship University of Russia (RUDN University), 
             ul. Miklukho-Maklaya 6, Moscow 117198, Russia}
\adr c {\small National Research Nuclear University ``MEPhI''
                    (Moscow Engineering Physics Institute), Moscow, Russia}
	}

%\Dates{October 31, 2018}
%{November 15, 2018}
%{November 20, 2018}

\Abstract
  {We discuss the peculiarities of gravitational lensing by spherically symmetric wormholes
   if they are not symmetric with respect to their throats. It is noticed, in particular, that 
   wormholes always contain the so-called photon spheres, near which the photon deflection 
   angles can be arbitrarily large, but, in general, the throat is such a sphere only for symmetric 
   wormholes. In some cases, photons from outside can cross the throat and return back 
   from a neighborhood of a photon sphere if the latter is located beyond the throat. Two 
   families of generally asymmetric wormhole configurations are considered as examples: 
   (1) anti-Fisher wormholes with a massless phantom scalar field as a source of gravity, and
   (2) wormholes with a zero Ricci scalar that may be interpreted as vacuum configurations 
   in a brane world. For these metrics, the photon effective potentials and deflection angles 
   are found and discussed.
}

] %%%%%%%%%%%%%%%%%%%%%%%%%%%%%%%%%%%%%5

\section{Introduction}

  Gravitational lensing is an astrophysical phenomenon whereby the propagation of light is
  affected by gravitating masses. As photons travel across the Universe, their trajectories 
  are  perturbed by gravitational fields. This phenomenon has become a powerful tool for 
  studying the distribution of mass in the Universe as well as for observing faint distant sources 
  that would otherwise be invisible.
  
  There is vast literature on light bending by different astrophysical bodies,
  beginning with Einstein's formula verified at solar eclipses (see, e.g., \cite{will}) and ending 
  with the most recent papers discussing gravitational lensing by strong-field objects like 
  \bhs\ and \whs, see, e.g., \cite{rev1,tsu-17} for reviews. However, in most of the papers, 
  only \whs\ symmetric with respect to their throats are considered, whereas such 
  objects are only a special subset among all \wh\ space-times. In this paper we discuss 
  some general features of gravitational lensing by asymmetric \whs, restricting 
  ourselves to \ssph\ configurations, \asflat\ on both sides from their throats. 
  
  We will consider two examples of such objects, one represented by the so-called 
  anti-Fisher solution to the Einstein equations with a 
  massless, minimally coupled phantom scalar field \cite{ber-lei, h_ellis, kb73},
  the other with an anisotropic fluid source and the Ricci scalar $R =0$ \cite{we-17}.
  In both cases, for simplicity, we assume that both the source of the signals and the 
  observer are located at very large distances from the deflecting body (the lens) as 
  compared to the characteristic lengths of the lens itself. Therefore, all incident paths 
  of photons are assumed to be initially parallel to each other. To calculate the 
  deflection angles, we use the corresponding general formulas which can be found, 
  e.g., in \cite{bozza-02, tsu-16}. A more general treatment taking into account finite 
  distances between all objects is presented in \cite{perlick}. Let us also mention 
  studies of gravitational lensing by rotating wormholes and the analysis of 
  specific features of photon paths in the strong deflection limit, see, e.g., 
  \cite{jusufi-17} and \cite{tsu-17}, respectively, as well as references given therein. 
  
  In Section 2 we discuss the general features of photon paths in asymmetric \whs,
  Section 3 is devoted to particular examples, and Section 4 is a conclusion.  

% ====================================================
\section{Photon spheres and lensing --- general consideration}

  The general static, spherically symmetric metric can be written as
\beq                     \label{ds1}
	ds^2=e^{2\gamma (u)}dt^2-e^{2\alpha (u)}du^2-e^{2\beta (u)}d\Omega ^2                     	
\eeq
  where $u$ is an arbitrary radial coordinate and $d\Omega=d\theta^2+\sin^2\theta d\phi^2$
  is the linear element on a unit sphere. Let us use the so-called quasiglobal 
  coordinate $u=x$ defined by the condition $\alpha +\gamma =0$, denoting 
\beq 		\label{qua}
	e^{2\gamma} = e^{-2\alpha} = A(x),\cm  e^{\beta}=r(x)	
\eeq
  so the metric takes the form
\beq                   \label{ds-q}
		ds^2 = A(x) dt^2  -\frac{dx^2}{A(x)} - r^2(x) d\Omega^2                                         	
\eeq
  It describes a \wh\ if both $A(x) >0$ and $r(x) >0$ for all $x \in \R$, and $r(x)$
  has a minimum (called a \wh\ throat) at certain $x$, say, $x=0$. To consider a \wh\
  as an object observable from distant weakly curved regions of space, we should 
  require that the metric is \asflat\ as $x \to \infty$: the same is required
  for $x \to -\infty$ if the other side of the \wh\ should be equally observable. In terms of 
  the metric \rf{ds-q} it means that
\bearr  				\label{asympt}
		A(x) = A_\pm + O(1/|x|), \quad\  r(x)\approx c_\pm\cdot |x|,
\nnn		 
		 Ar'{}^2 \to 1  \quad\ {\rm as} \ \  x\to \pm\infty,
\ear
   where $A_\pm, c_\pm $ are positive constants, and the prime stands for $d/dx$.
        
  For the metric \rf{ds-q}, the null geodesic equations (assuming photon motion in the 
  equatorial plane $\theta=\pi/2$ without losing generality) has two well-known integrals
\beq  			\label{geo-int}
	{\dot t} = E/A(x), \qquad      {|\dot\phi|} = L/r^2(x),
\eeq    
  where dots denote $d/d\sigma$,  $\sigma$ being an affine parameter along the 
  geodesics; $E$ is the conserved energy parameter, and $L$ the conserved angular
  momentum. With these integrals, the condition $u^\mu u_\mu =0$ for the photon
  4-velocity $u^\mu$ (comprising an integral of the radial component of the 
  null geodesic equations) reads
\beq                        \label{geo-r}
  		{\dot x}^2 + L^2 A(x)/r^2(x) = E^2.
\eeq  		
  This is the energy conservation law at photon motion: the first term plays the role of 
  kinetic energy while the second one that of a potential, $V(x) = L^2 A/r^2$.
  The photon motion is only possible in the range of $x$ where $E \geq V(x)$. In particular,
  circular photon orbits ($\dot x \equiv 0$) can take place on such spheres $x = x_{\rm ph}$ 
  where $V' =0$, that is, according to \rf{geo-r},
\beq  								\label{circ}
             A/r^2 = E^2/L^2,  \qquad   r A' - 2A r' =0.               
\eeq   
  
  Spheres \rf{circ} (the so-called {\it photon spheres\/}) play an important role in 
  gravitational lensing since the deflection angle of a photon approaching such a sphere 
  along its tangent tends to infinity. It is therefore of interest to compare the 
  location of photon spheres in different space-times. 
  
  Thus, in a \Swz\ \bh\  space-time ($r=x$, $A = 1 - 2m/r$) with mass $m$, the 
  photon sphere is located at $ r = 3m$ ($3/2$ of the horizon radius). 
  
%\red{RN: we'll maybe discuss}
  
  For \wh\ space-times, the \Swz\ mass, describing the metric at large $|x|$ (assuming
  that it is \asflat\ on both sides of the throat), is in general not the only parameter 
  determining their shape: there is at least one more characteristic length, the throat radius 
  $r_{\rm th} = \min r(x)$. Furthermore, the \wh\ may be symmetric or asymmetric with 
  respect to the throat, which, without loss of generality, can be placed at $x=0$.  
  If the \wh\ is symmetric, both $r(x)$ and $A(x)$ are even functions, hence
  $r' = A' =0$ at the throat, and by \rf{circ} the throat is inevitably a photon sphere.   
  (This is also evident by symmetry: a photon launched from a point on the throat in one 
  of the angular directions, that is, with $\dot x =0$, has no reason to leave the sphere 
  $x=0$ to either positive or negative $x$.) A symmetric \wh\ may have, but not necessarily 
  has, other photon spheres, depending on the behavior of the metric functions $A$ and 
  $r$, as was recently discussed in \cite{shaikh-18}. It is necessary to note that the claim 
  of  \cite{shaikh-18} that a throat is necessarily a photon sphere is true for symmetric 
  \whs\ only.
  
  If the \wh\ is asymmetric, such that $A'(0) \ne 0$, then by \rf{circ} the throat $x=0$
  is not a photon sphere. On the other hand, since in a twice \asflat\ \wh\ $V(x) \to 0$ 
  at both infinities, it is clear that there should be at least one photon sphere as a 
  maximum of $V(x)$, but the actual number and allocation of photon spheres depends 
  on the particular metric. One should also note that photon paths along such
  a sphere are stable if $V(x)$ has there a minimum and are unstable otherwise.
  
  In what follows we will find the positions of photon spheres and the corresponding 
  features of gravitational lensing for some examples of \wh\ metrics.

  To describe gravitational lensing, one can use the general formulas for
  asymptotically flat static, spherically symmetric space-times, see, e.g., 
  \cite{bozza-02, tsu-16}. We will assume that both the source and the 
  observer are located at very large distances from the deflecting body (the lens) as 
  compared to the characteristic lengths of the lens itself. Then all 
  incident paths of photons are initially parallel to each other.  
  
  In this case, in our notations, the deflection angle $\alpha$, found from null 
  geodesics in the metric \rf{ds-q}, is given by
\bearr
	\alpha = \alpha(x_0) = I(x_0) - \pi,	
\nnn								\label{lens1}
          I(x_0) = 2 \int_{x_0}^\infty \frac {dx}{r(x) \sqrt {r^2(x)/b^2 - A(x)} },
\ear 
  where $b = L/E$ is the impact parameter characterizing a particular null geodesic 
  with the conserved energy parameter $E$ and angular momentum $L$ according to 
  \rf{geo-int}; $x_0$ is the coordinate value corresponding to the nearest approach 
  of the photon to the strong field region and can be found from the condition 
  $dr/d\sigma =0$ which leads to
\beq 							\label{lens2}
	A(x_0) b^2 = r^2 (x_0),
\eeq
  and $I(x_0)$ in \eqn{lens1} turns into  
\beq                             				 \label{lens3}
          I(x_0) = 2 \int_{x_0}^\infty \frac {r_0 dx}{r(x) \sqrt {r^2(x)A(x_0) - r_0^2 A(x)} },
\eeq          
  where $r_0 = r(x_0)$.
  Equations \rf{lens1} and \rf{lens3} will be further used for finding specific \wh\ lensing
  characteristics.
  
% ==================================
\section{Examples}
% -------------------------------------------------------------
\subsection{Lensing by anti-Fisher \whs}
    
   One of the simplest known \wh\ solutions in \GR\ is the so-called anti-Fisher solution 
   to the Einstein-minimally coupled massless scalar field system in the case where the 
   scalar field $\phi$ is phantom, and its Lagrangian is\footnote 
	   {The term ``anti-Fisher'' has been suggested in order to recall that the corresponding
	    solution for a canonical scalar field was first obtained by I.Z. Fisher in 1948 \cite{fisher}. }
   $L_\phi  = - g\MN \phi_{,\mu}\phi_{,\nu}$ \cite{ber-lei, h_ellis, kb73}.  
   The \wh\ branch of the anti-Fisher solution may be written in the form \cite{kb73, br-book}
\bearr                                                                                          \label{ds2}
		ds^2 = e^{-2mu} dt^2 - \frac{k^2 e^{2mu}}{\sin^2 ku} 
					\bigg[ \frac{k^2 du^2}{\sin^2 ku} + d\Omega^2 \bigg],
\nnn
	        \phi = Cu, \cm     k^2 + m^2 = C^2/2,
\ear	       
   where $k>0$, $m$ and $C$ are integration constants, and $C$ has the meaning of a 
   scalar charge. The radial coordinate $u$ is harmonic: it satisfies the ``gauge''
   condition $\alpha = 2\beta + \gamma$ in terms of the metric \rf{ds1}, which implies
   $\Box u =0$. Without loss of generality we assume $0 < u < \pi/k$, and the metric 
   \rf{ds2} is \asflat\ at both ends of this range, where the spherical radius $r = e^\beta$ 
   is infinite. The constant $m$ is the \Swz\ mass at spatial infinity $u=0$, as verified by 
   comparison with the \Swz\ metric at small $u$. At the other spatial infinity,
   $u = \pi/k$, the \Swz\ mass is $m_- = -m e^{-m\pi/k}$. Thus, if the mass is positive 
   at one end, $m >0$, it is negative on the other end and larger than $m$ in absolute 
   value.\footnote
   	{This inequality was discussed by H. Ellis in his recent paper \cite{ellis-15} 
   	  as a possible source of dark energy due to net negative mass of such \whs.}
   
   The solution \rf{ds2} is transformed to the quasiglobal coordinate $x$ (which is 
   approximately equal to the conventional radial coordinate at large radii) by the substitution   
\beq    					\label{u->x}
		ku = \arccot (x/k),\qquad   x = k \cot ku,  
\eeq          
   whence (preserving the notation \rf{u->x} for $u$)
\beq  					\label{ds3}	
		ds^2 =e^{-2m u} dt^2 - e^{2mu}[dx^2 + (k^2 + x^2) d\Omega^2].
\eeq      
   The two flat spatial infinities correspond to $x \to \pm \infty$. The \wh\ throat, where $r$
   has its minimum, is located at  $x = x_{\rm th} = m$ and has the size
\beq
                r_{\rm th} = \sqrt{m^2 + k^2} \exp \Big( \frac mk \arccot \frac mk \Big).
\eeq      
   The photon sphere parameters found using \rf{circ} are
\bearr
                x_{\rm ph} = 2m, 
\nnn
	                r_{\rm ph} = \sqrt{4m^2 + k^2} \exp \Big( \frac mk \arccot \frac {2m}k \Big).
\ear    
   The photon sphere coincides with the throat only in the massless case $m=0$, often
   called the Ellis \wh, although in fact H. Ellis in  in \cite{h_ellis} considered the general 
   anti-Fisher solution.
   
   For numerical estimates of light deflection angles by an anti-Fisher \wh, we put
   $k =1$, thus specifying an arbitrary length scale, and the parameter $m$ also
   becomes dimensionless. Our length unit is then equal to the throat radius $r_{\rm th}$ 
   for $m=0$ but rather strongly differs from it for $m \ne 0$, see Fig.\,1. The same 
   figure also shows the photon sphere radius $r_{\rm ph}$: its difference from $r_{\rm th}$
   looks rather small, apparently because $r(x)$ is a slowly varying function close
   to its minimum.
% -----------------------------------------
\Figu{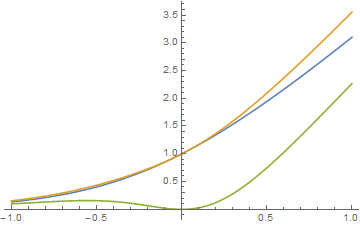}{7cm}
	{Parameters of the anti-Fisher solution with $k=1$. 
	Upside-down: radius of the photon sphere $r_{\rm ph}(m)$, throat radius  $r_{\rm th}(m)$,
	and their difference multiplied by 5 for better visibility.}
% ------------------------------------------	
     
   For the metric \rf{ds3} with $k=1$, the integral \rf{lens3} reads
\bearr  
	I(x_0) = \int\limits_{x_0}^\infty \ \frac{2 \sqrt{(1+x_0^2)/(1+x^2)}\ dx}
				{\sqrt{(1+x^2) e^{4m(a_0-a)} - (1+x_0^2)}},
\nnnv
	a := \arctan x, \qquad a_0 := \arctan x_0.				
\ear
  This expression can be used to calculate the deflection angles $\alpha(x_0) = I(x_0) - \pi$
  for $x_0 > 2m$, and the results are presented in Fig.\,2. 
% -----------------------------------------
\Figu{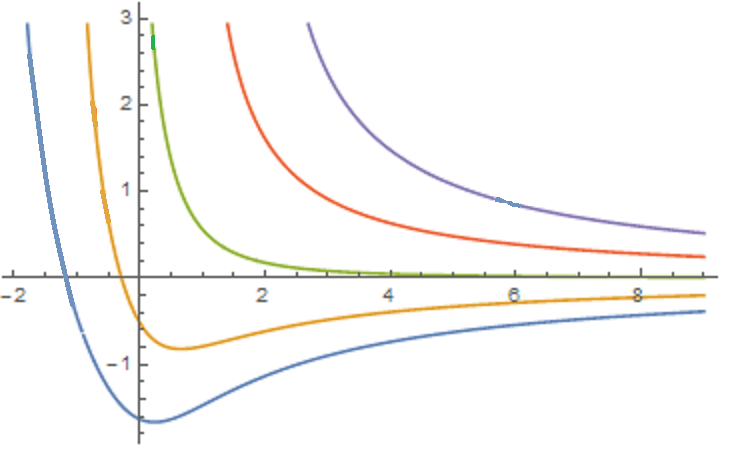}{7cm}
	{Deflection angles $\alpha$ as functions of the closest approach coordinate $x_0$ 
	for anti-Fisher \whs\ with $k=1$ and masses $m = 1,\ 0.5,\ 0,\ -0.5,\ -1$ (upside-down). }
% ------------------------------------------	      
  The calculation confirms that the deflection angles diverge near the photon spheres 
  $x_{\rm ph} = 2m$. In the massless case (Ellis \wh) the deflections are especially 
  small except for the immediate neighborhood of the throat. As could be expected, 
  negative masses create negative deflections (out from the lens), but close to the  
  photon sphere they anyway become positive. To approach this sphere, located 
  beyond the throat, photons have also to travel partly beyond the throat and only then 
  return to ``our'' side of the \wh.
  
   With $m >0$, the nature of deflections is more or less conventional. What can be marked 
   as a general feature of \whs\ is the photon sphere location close to $r=m$, whereas
   for a \Swz\ \bh\ its radius is $r_{\rm ph} = 3m/2$. It is an evident potentially observable 
   distinction.  
   
% -----------------------------------------------------------------   
\subsection{Lensing by some \whs\ with $R=0$}   

  Symmetric and asymmetric \whs\ with anisotropic fluid sources were obtained in 
  \cite{we-17} among other solutions to the Einstein equations under the assumptions 
\beq\nhq
		R = 0, \quad r(x) = \sqrt{x^2 + b^2}, \quad   b = \const > 0,      \label{R-sol}      
\eeq    
   where $R$ is the Ricci scalar, and the metric is assumed in the form \rf{ds-q}. 
   The condition $R=0$ allows for treating such solutions as those describing either
   the gravitational field of an anisotropic fluid source or as vacuum configurations
   in an RS2-type brane world, like those considered in \cite{br-kim}, since $R=0$ 
   follows from the brane-world modified 4D Einstein equations \cite{BW-SMS}
   in vacuum. 
   
   According to \rf{R-sol}, in such solutions the \wh\ throat is located at $x=0$, and its radius
   is $r_{\rm th} =b$. Assuming $b=1$ for numerical calculations, we choose the throat 
   radius as the length unit.  
   
   As examples, we take solutions to the equation $R(x) =0$, having the form 
\beq                                    \label{eq-R0}
	  A'' + \frac{4x}{1+x^2} A' + \frac{2(2+x^2)}{(1+x^2)^2} A = \frac{2}{1+x^2},
\eeq    
  under the initial conditions $A(0) = 0.5$ and $A'(0) = 0, \pm 0.5$. 
  The solutions are obtained numerically.  With $A'(0) =0$
  we obtain an even function $A(x)$, corresponding to a symmetric \wh, while those
  obtained with $A'(0) = \pm 0.5$ describe significantly asymmetric \whs, see Fig.\,3.
  Fig.\,4 shows the behavior of the effective potential for photon motion. The positions 
  of its maximum values show the location of photon spheres: 
  $x_{\rm ph} = 0$ and $r_{\rm ph} =1$ for the symmetric solution ($A'(0) =0$), 
  $x_{\rm ph} \approx \pm 0.3031$ and $r_{\rm ph} \approx 1.0445$ for the asymmetric 
  solutions ($A'(0) = \pm 0.5$). 
% ----------------------------------------- fig 3
\Figu{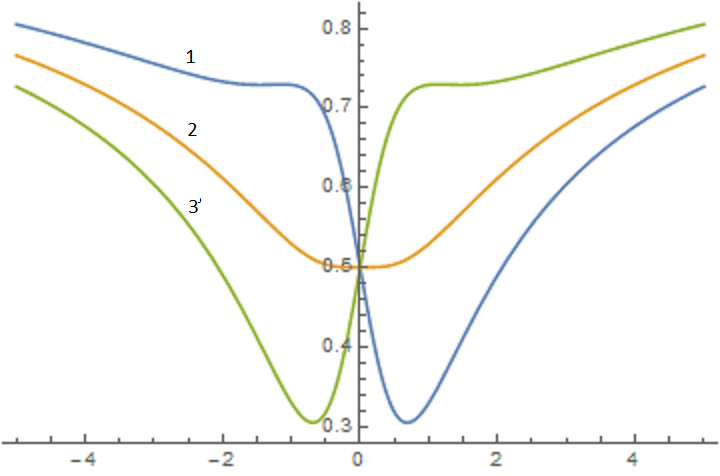}{7cm}
	{The metric function $A(x)$ as a solution to \eqn{eq-R0} with $A(0) = 0.5$ and 
	 $A'(0) = -0.5,\ 0,\ 0.5$ (curves 1, 2, 3, respectively).}
% ------------------------------------------ fig 4
\Figu{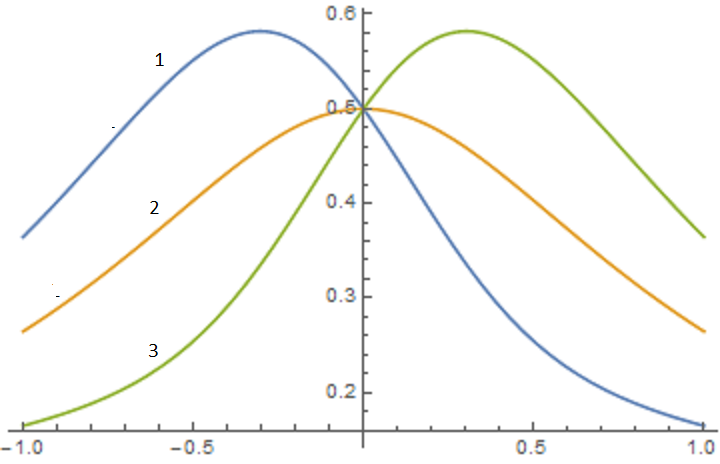}{6.5cm}
	{The quantity $V(x) = A(x)/r^2(x)$ for the solutions shown in Fig.\,3; the maxima 
	  of $V(x)$ show the positions of photon spheres.}
% ------------------------------------------ 
   
   Figure 5 shows the deflection angles $\alpha(x_0)$ for these three examples of \wh\
   configurations. The plots confirm that the deflection blows up if a photon approaches
   a photon sphere, but the behavior of $\alpha(x_0)$ contains some peculiar features. 
   Thus, a ``zigzag'' of curve 1 at small positive $x$ is evidently related to a quick rise of
   $A(x)$ while approaching the throat ($x=0$) from the positive side. One can also 
   observe that the radii of the photon spheres are rather close to the throat radius, at 
   least if the \wh\ asymmetry is not too large.
% ------------------------------------------ fig 5
\Figu{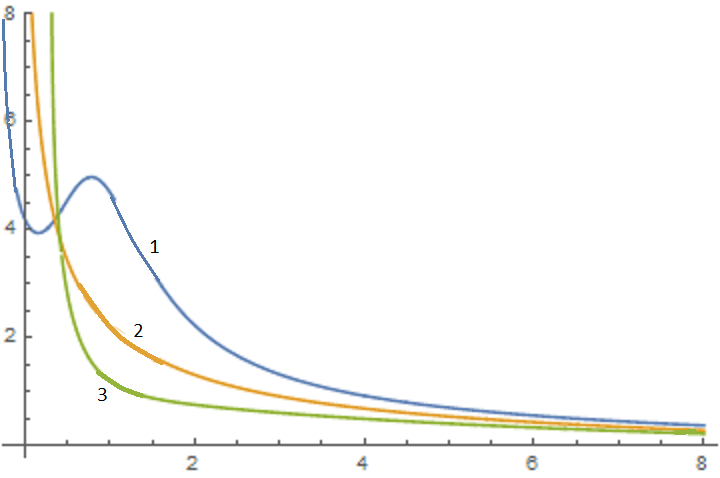}{7.5cm}
	{Deflection angles $\alpha$ as functions of the closest approach coordinate $x_0$ 
	for \whs\ with $R=0$ and $A(x)$ shown in Fig.\,3 (the curves are numbered accordingly).}  
% ------------------------------------------     
     
% =============================
\section{Concluding remarks}

   We have discussed some features of gravitational lensing by asymmetric \whs\ and 
   can make the following observations.
   
   {\bf 1.} All \asflat\ \ssph\ \whs\ possess photon spheres near which there can be arbitrarily
   large deflection angles of incident photons.
   
   {\bf 2.} A photon sphere is necessarily located on the throat if the \wh\ is symmetric with 
   respect to it, while in generic asymmetric \whs\ the throat is, in general, not a photon sphere.
   
   {\bf 3.} If a photon sphere is located beyond the throat, some photons can also travel 
   beyond the throat and return back into the initial region.
   
   {\bf 4.} The deflection angle dependence on the closest approach coordinate $x_0$ 
   is not always monotonic and depends on the particular \wh\ geometry.
   
   {\bf 5.} We confirm the already known observation that the location of photon spheres is
   not directly related to the \Swz\ mass of a \wh; the latter can even be massless. However,
   in known examples the size of a photon sphere is smaller than that of a \Swz\ \bh\ of the 
   same mass, which may be an observational distinction between black holes and \whs.  
   
   As particular examples, we have considered anti-Fisher wormholes and those with a zero 
   Ricci scalar that may be interpreted as vacuum configurations in a brane world. 
   One can recall that anti-Fisher \whs\ are unstable under radial perturbations 
   \cite{gonz, we-11, we-12} and therefore cannot be regarded viable. However, we used them 
   here as convenient examples of generally asymmetric \whs\ (symmetric only in the massless 
   case $m=0$), convenient for studying the asymmetry effects in gravitational lensing. One can 
   also recall that the same metric can be obtained with another source of gravity having the 
   same stress-energy tensor in the static case but quite different dynamic properties, including 
   stability, as exemplified in \cite{br-shaz} for Ellis \whs.
                
   In future studies it makes sense, among other things, to clarify the photon behavior close to 
   the photon spheres in generic asymmetric \whs, to take into account finite distances between 
   the \wh, the source and the observer, affecting the observable picture in the sky,
   and to describe the gravitational lensing picture in the case of multiple photon spheres
   (see, e.g., \cite{shaikh-18}). It would also be of interest to consider the effect of long \wh\ 
   throats \cite{long-17} on gravitational lensing and to extend the study to other kinds 
   of \whs\ such as those known in scalar-tensor and $f(R)$ theories of gravity (see, e.g. 
   \cite{kb73, kb-star, kb18}) and those with ``trapped ghosts'' where a scalar field is phantom
   only close to the throat  \cite{kb-sush, kb18}.    
   
\subsection*{Acknowledgments}

  This publication has been prepared with the support the RUDN University Program 5-100.
  The work of KB was also partly performed within the framework of the Center FRPP supported 
  by MEPhI Academic Excellence Project (contract No. 02.a03. 21.0005, 27.08.2013).

\end{document}

%% file: preamble.tex
\usepackage{amsfonts,amssymb,cite}
\usepackage{graphicx}

\def\noi{\noindent}

\newcommand{\Title}[1]{\noi {{\Large\bf #1}}\\[1ex]}

\def\Aunames#1{\noi{\bf #1}}
\def\au#1{${}^{#1}$}
\def\Addresses#1{\medskip\noi \protect
	\begin{description}\itemsep -3pt {\it #1} \end{description}}
\def\adr#1#2{\item[${}^{#1}$]{\it #2}}

\newcommand{\Abstract}[1]{\vskip 2mm \begin{center}
        \parbox{16.4cm}{\small\noi #1} \end{center}\medskip}

\def\nqq{\hspace*{-2em}}
\def\nhq{\hspace*{-0.5em}}

\def\cm{\hspace*{1cm}}

\def\Jl#1#2{#1 {\bf #2},\ }

\def\ApJ#1 {\Jl{Astroph. J.}{#1}}
\def\CQG#1 {\Jl{Class. Quantum Grav.}{#1}}
\def\DAN#1 {\Jl{Dokl. AN SSSR}{#1}}
\def\GC#1 {\Jl{Grav. Cosmol.}{#1}}
\def\GRG#1 {\Jl{Gen. Rel. Grav.}{#1}}
\def\JETF#1 {\Jl{Zh. Eksp. Teor. Fiz.}{#1}}
\def\JETP#1 {\Jl{Sov. Phys. JETP}{#1}}
\def\JHEP#1 {\Jl{JHEP}{#1}}
\def\JMP#1 {\Jl{J. Math. Phys.}{#1}}
\def\NPB#1 {\Jl{Nucl. Phys. B}{#1}}
\def\NP#1 {\Jl{Nucl. Phys.}{#1}}
\def\PLA#1 {\Jl{Phys. Lett. A}{#1}}
\def\PLB#1 {\Jl{Phys. Lett. B}{#1}}
\def\PRD#1 {\Jl{Phys. Rev. D}{#1}}
\def\PRL#1 {\Jl{Phys. Rev. Lett.}{#1}}

\def\lal{&&\nqq {}}

\def\beq{\begin{equation}}
\def\eeq{\end{equation}}
\def\bear{\begin{eqnarray}}
\def\bearr{\begin{eqnarray} \lal}
\def\ear{\end{eqnarray}}
\def\earn{\nonumber \end{eqnarray}}

\def\nnn{\nonumber\\ \lal }
\def\nnnv{\nonumber\\[5pt] \lal }

\def\const{{\rm const}}

%% file: lens-b.bbl
\small
\begin{thebibliography}{99}  \itemsep .5pt
     
\bibitem{will}
	C. M. Will, ``The confrontation between general relativity and experiment'', 
	Living Rev. Rel. {\bf 17}, 4 (2014); arXiv: 1403.7377.

\bibitem{rev1}
	Hideki Asada, Gravitational lensing by exotic objects.
	Mod. Phys. Lett. A {\bf 32}, 1730031 (2017); arXiv: 1711.01730.

\bibitem{tsu-17}	
	Naoki Tsukamoto,
	Deflection angle in the strong deflection limit in a general asymptotically flat, 
	static, spherically symmetric spacetime.
	Phys. Rev. D {\bf 95}, 064035 (2017); arXiv: 1612.08251.

\bibitem{ber-lei} 
	O. Bergmann and R. Leipnik,
	Space-time structure of a static spherically symmetric scalar field.
        Phys. Rev. {\bf 107}, 1157 (1957).
	
\bibitem{h_ellis} 
	Homer G. Ellis, 
	Ether flow through a drainhole: A particle model in general relativity,
	J. Math. Phys. {\bf 14}, 104 (1973).
	
\bibitem{kb73}
	K. A. Bronnikov, Scalar-tensor theory and scalar charge,
	Acta Phys. Pol. B {\bf 4}, 251 (1973).
	
\bibitem{we-17}
	K. A. Bronnikov, K. A. Baleevskikh, and M. V. Skvortsova,
	Wormholes with fluid sources: A no-go theorem and new examples.
	\PRD {96} 124039 (2017); arXiv: 1708.02324.
	
\bibitem{bozza-02}
	V. Bozza, Gravitational lensing in the strong field limit,
	Phys. Rev. D {\bf 66},  103001 (2002).

\bibitem{tsu-16}
	Naoki Tsukamoto,
	Strong deflection limit analysis and gravitational lensing of an Ellis wormhole
	Phys. Rev. D 94, 124001 (2016); arXiv: 1607.07022

\bibitem{perlick}
	Volker Perlick,
	On the exact gravitational lens equation in spherically symmetric and static spacetimes.
	Phys. Rev. D {\bf 69}, 064017 (2004); gr-qc/0307072.
	
\bibitem{jusufi-17} 
	Kimet Jusufi and Ali \"Ovg\"un,
	Gravitational lensing by rotating wormholes.
	Phys. Rev. D {\bf 97}, 024042 (2018); arXiv: 1708.06725 

\bibitem{shaikh-18}
	Rajibul Shaikh, Pritam Banerjee, Suvankar Paul, and Tapobrata Sarkar,
	A novel gravitational lensing feature by wormholes.
	arXiv: 1811.08245.

\bibitem{fisher}
	I.Z. Fisher, Scalar mesostatic field with regard for gravitational effects,
	Zh. Eksp. Teor. Fiz. {\bf 18}, 636 (1948); gr-qc/9911008.

\bibitem{br-book}
	K. A. Bronnikov and S. G. Rubin, 
	{\it Black Holes, Cosmology, and Extra Dimensions} (World Scientific, Singapore, 2012).

\bibitem{ellis-15}
	Homer G. Ellis.
	Cosmology without Einstein's assumption that inertial mass produces gravity
	Int. J. Mod. Phys. D {\bf 24}, 1550069 (2015); gr-qc/0701012.

\bibitem{gonz}
	J. A. Gonzalez, F. S. Guzman, and O. Sarbach, 
	Instability of wormholes supported by a ghost scalar field. 
	I. Linear stability analysis. Class. Quantum Grav. {\bf 26}, 015010 (2009).

\bibitem{we-11}
	K. A. Bronnikov, J. C. Fabris, and  A. Zhidenko,
	On the stability of scalar-vacuum space-times,
	Euro Phys. J. C {\bf 71}, 1791 (2011); ArXiv: 1109.6576.
	
\bibitem{we-12}	
	K. A. Bronnikov, R. Konoplya, and A. Zhidenko,
	Instabilities of wormholes and regular black holes supported by a phantom scalar field.
	Phys. Rev. D {\bf 86}, 024028 (2012); arxiv: 1205.2224.

\bibitem{br-shaz}
	K. A. Bronnikov, L. N. Lipatova,. I. D. Novikov, and A. A. Shatskiy,
	Example of a stable wormhole in general relativity,
	\GC {19} 269 (2013); arXiv: 1312.6929

\bibitem{br-kim}
	K. A. Bronnikov and S.-W. Kim, Possible wormholes in a brane world, 
	Phys. Rev. D {\bf 67}, 064027 (2003).

\bibitem{BW-SMS}
	T. Shiromizu, K. Maeda, and M. Sasaki, 
	The Einstein equations on the 3-brane world, Phys. Rev. D {\bf 62}, 024012 (2000).

\bibitem{long-17}
	K. A. Bronnikov and P. A. Korolyov, 
	On wormholeas with long throats and the stability problem.
        Grav. Cosmol. {\bf 23}, 273 (2017); arXiv: 1705.05906.

\bibitem{kb-star}
	K. A. Bronnikov, M. V.  Skvortsova, and A. A. Starobinsky, Notes on wormhole existence 
	in scalar-tensor and $F(R)$ gravity, Grav. Cosmol. {\bf 16}, 216 (2010); arXiv: 1005.3262.

\bibitem{kb18}
	K. A. Bronnikov, Scalar fields as sources for wormholes and regular black holes,
	Particles {\bf 1}, 5 (2018); arXiv: 1802.00098.
	
\bibitem{kb-sush}
	K. A. Bronnikov and S. V. Sushkov, Trapped ghosts: A new class of wormholes. 
	Class. Quantum Gravity {\bf 27}, 095022 (2010); arXiv: 1001.3511.

\end{thebibliography}
